\documentclass{article}
\usepackage{authblk}       % creates author & affiliation blocks
\usepackage{amsmath, amssymb}
\usepackage{xfrac}           % allows slanted fractions \sfrac{num}{denom}
\usepackage{fullpage}       % one inch margins
\usepackage{graphicx}      % should permit using .pdf or .ps graphics
\usepackage{placeins}        % allows \FloatBarrier at end of sections
\usepackage{parskip}          % gives spaced nonindented paragraphs.
\usepackage[T1]{fontenc}      % allows use use "<" and ">" directly, also vertical bar "|" 
\usepackage[font=footnotesize]{caption}   % just right if body text is 'small'
\usepackage{float}                  % allows MUCH better control of float placement [H]

\providecommand{\keywords}[1]
{
  \textbf{\textit{Keywords---}} #1
}

\begin{document}
\let\footnotesize\small
\date{}      % empty so no date showing
\title{Doomsday: A Response to Simpson's Second Question}
\author[]{Mike Lampton}
\affil[]{Space Sciences Lab, UC Berkeley}
\affil[]{\textit{mlampton@berkeley.edu}}

\maketitle
\large
\begin{abstract}  
\noindent
{The Doomsday Argument (DA) has sparked a variety of opinions.  Here I address a key 
question posed by F. Simpson (2016) that confronts the views of DA proponents
and those who, like me, oppose the DA.  I agree that typical locations within
a complete \textit{spatial} distribution are calculable using ordinary frequentist probability.
But I argue that the \textit{temporal} probability distribution is unknown: we have records of our past
yet are ignorant of our future.  It is this asymmetry that upsets the idea of Copernicanism
in time.  Although frequentist methods do not apply to this asymmetric situation, Bayesian methods do apply.
They show that the various Quick Doom and Distant Doom scenarios are equally likely.
I conclude that the DA has no predictive power whatsoever.}
\end{abstract}

\begin{center}
\end{center}

\large
\section{Introduction}   % section 1

How long into our future will mankind last?   The  ``Doomsday Argument'' (Carter 1983, Gott 1993, 
Leslie 1996, Leslie 2010)   is an effort to answer this question quantitatively.  The argument 
is based on the fact that samples from a known distribution
are rarely found at its extreme ends.  In only 1\% of random draws would a sample be found in
the distribution's  first percentile zone.  So, the argument goes, there is only a 1\% chance of finding
our present human rank number less than $1 \times 10^{11}$  if mankind's total span is $1 \times 10^{13}$  
individuals.  Larger spans make this probability even smaller.  Since we are all reluctant to accept unlikely 
situations, we should be reluctant to accept the notion that the total span of humanity will exceed  
$1 \times 10^{13}$ individuals. 

I will not attempt to summarize the many discussion points on this topic except to direct the reader to
five related discussions (Oliver and Korb 1998, Monton and Roush 2001, Northcott 2016,  McCutcheon 2018,
 Lampton 2019).  Here I want to focus on question 2 raised by Fergus Simpson (2016) that 
 concisely captures the issue: can probabilistic arguments constrain our future?

\begin{quote}
``Q2. Why would your spatial location relative to other humans appear 
representative of the parent population, but not your temporal one?''
\end{quote}

\newpage
\section{Spatial Distributions}  % section 2
In Figure 1, I show the distribution of the world population with respect to latitude.
Living in California, I find myself at latitude $+37\deg$ and so I am at about the 60th percentile
of latitudes, typically situated among the people of the world: not an extreme outlier. 
The entire distribution is known: each individual can be located in
the chart, according to latitude, with the span $-90\deg$ to $+90\deg$ encompassing the
entire human race.   Because the distribution is known, frequentist probability applies.
I must find myself in one of these one-degree bins, so the sum of my bin probabilities is 1. 

\vspace{6mm}

\begin{figure}[ht] 
\begin{center}
     \includegraphics{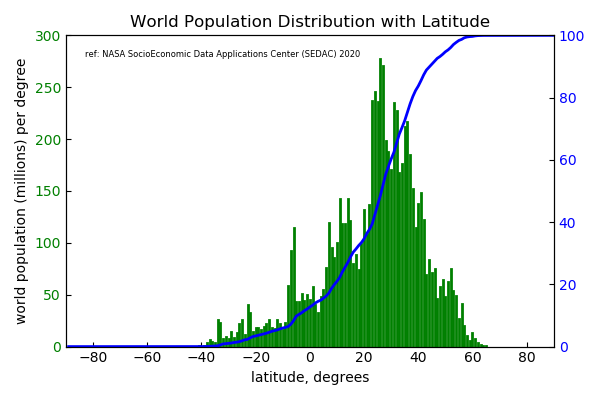}  % Fig 1
     \caption{World population vs latitude. Green: differential. Blue: cumulative percent.  }
 \end{center}
\end{figure}

\section{Temporal Distributions}  % section 3
In Figure 2, I show the cumulative growth of world population versus time.  There is
a key difference from the cumulative distribution seen in Figure 1: here it is plotted as the
number of individuals, not as a percentage.  It is merely the beginning of an integral,
and unlike Figure 1, it cannot be normalized to $100\%$ because we do not have the
future data required to normalize it.  

\newpage
\begin{figure}[ht] 
\begin{center}
     \includegraphics{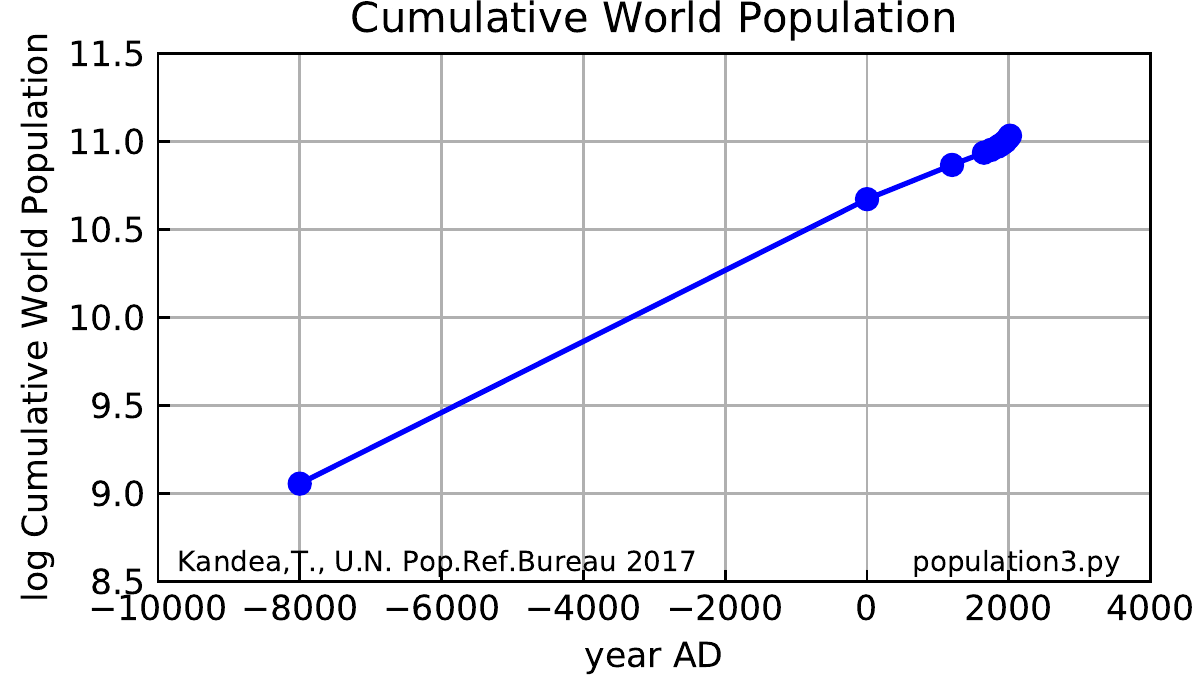}  % Fig 2
     \caption{Cumulative number of persons versus time, worldwide}
 \end{center}
\end{figure}

\section{Conclusion}
Returning to Simpson's question 2, how do these cases differ?  My reply is that
the spatial distribution is known, and my rank within this distribution is in principle 
measurable.  In contrast, the eventual extent of mankind's growth 
is unknown: we can gather population data from our past but not from our future.  
Given some evidence, Bayesian logic provides a way to quantitatively compare two 
alternative hypotheses.   In an earlier treatment (Lampton 2019), I evaluated the
relative likelihoods of various Quick Doom and Distant Doom scenarios given the
evidence that mankind has, so far, numbered about $1 \times 10^{11}$ individuals.
The result is exactly what you would expect:  given only past data, those scenarios 
are equally likely.  Given only what we know now, we have no way to decide between them.

\section*{References}

Carter, B. "The anthropic principle and its implications for biological evolution," Phil. Trans. Royal Soc. 
London A310, pp.347-363 (1983). 

Gott, J.R., "Implications of the Copernican Principle for our future prospects," Nature v.363, 27 (1993). 

Lampton, M., "Doomsday: Two Flaws," arXiv 1909.11031 (2019).

Leslie, J., "The End of the World," London and New York: Routledge, (1996).  

Leslie, J., "Risk that Humans will soon be extinct,"  Philosophy v85 No.4 (2010). 

McCutcheon, R., "What, Precisely, is Carter's Doomsday Argument?" https://philpapers.org, (2018)

McCutcheon, R., "In Favor of Logarithmic Scoring," Phil. Sci. v86 No.32  286-303 (2019)

Monton B. and Roush S., "Gott's Doomsday Argument,"  https://philpapers.org (2001).

Northcott, R., "A Dilemma for the Doomsday Argument," https://philpapers.org (2016).

Oliver, J. J., and Korb,  "A Bayesian Analysis of the Doomsday Argument,"  Monash U. (1998).

Simpson, F., "Apocalypse Now? Reviving the Doomsday Argument," arXiv 1611.03072 (2016).

\end{document}